\documentclass[journal = jctcee, manuscript = article, layout = twocolumn]{achemso}
\usepackage{natbib}
\usepackage{amsmath}
\usepackage{bm}
\usepackage{graphicx}
\usepackage{hyperref}
\usepackage{xcolor}

\title{Is the Trotterized UCCSD Ansatz Chemically Well-Defined?}

\author{Harper R. Grimsley}
\affiliation{Department of Chemistry, Virginia Tech, Blacksburg, VA 24061, USA}
\author{Daniel Claudino}
\affiliation{Department of Chemistry, Virginia Tech, Blacksburg, VA 24061, USA}
\author{Sophia E. Economou}
\affiliation{Department of Physics, Virginia Tech, Blacksburg, VA 24061, USA}
\author{Edwin Barnes}
\affiliation{Department of Physics, Virginia Tech, Blacksburg, VA 24061, USA}
\author{Nicholas J. Mayhall}
\email{nmayhall@vt.edu}
\affiliation{Department of Chemistry, Virginia Tech, Blacksburg, VA 24061, USA}

\begin{document}

\begin{abstract}
The variational quantum eigensolver (VQE) has emerged as one of the most promising near-term quantum algorithms that can be used to simulate many-body systems such as molecular electronic structures. Serving as an attractive ansatz in the VQE algorithm, unitary coupled cluster (UCC) theory has seen a renewed interest in recent literature. However, unlike the original classical UCC theory, implementation on a quantum computer requires a finite-order Suzuki-Trotter decomposition to separate the exponentials of the large sum of Pauli operators.  While previous literature has recognized the non-uniqueness of different orderings of the operators in the Trotterized form of UCC methods, the question of whether or not different orderings matter at the \textit{chemical scale} has not been addressed. 
In this letter, we explore the effect of operator ordering on the Trotterized UCCSD ansatz, as well as the much more compact $k$-UpCCGSD ansatz recently proposed by Lee et al. [J. Chem. Theory Comput., \textbf{2019}, \textit{15}, 311; arXiv, \textbf{2019}, quant-ph:1909.09114] We observe a significant, system-dependent variation in the energies of Trotterizations with different operator orderings. The energy variations occur on a chemical scale, sometimes on the order of hundreds of kcal/mol. This letter establishes the need to define not only the operators present in the ansatz, but also the order in which they appear. This is necessary for adhering to the quantum chemical notion of a ``model chemistry'', in addition to the general importance of scientific reproducibility. As a final note, we suggest a useful strategy to select out of the combinatorial number of possibilities, a single well-defined and effective ordering of the operators.
\end{abstract}

\maketitle
\section{Introduction}
The ability to accurately simulate chemistry at the sub-atomic level can provide deeper scientific insights and further reaching predictions than through experiment alone. 
Although exact simulation requires computational resources which increase exponentially with system size,
    many stable molecules can be accurately modeled using polynomially scaling techniques, providing accurate and interpretable results.
Examples of such approximations include density-functional theory, perturbation theory, or coupled-cluster theory.
To study more complicated systems with many strongly correlated electrons such as those involved in numerous catalytic systems or materials applications, more general modeling solutions are needed. 

Quantum simulation, which has recently seen a dramatic increase in activity due to rapid developments in both hardware and algorithms, 
provides an exciting possibility for performing approximation-free simulations without the exponential computational cost plaguing classical simulations. Because the Hilbert space of a single spin-orbital can be mapped to the Hilbert space of a single qubit, the exponential growth of the molecular Hamiltonian is matched by the exponential growth of a quantum computer's Hilbert space. Consequently, a quantum computer with only tens of logical qubits could potentially demonstrate a quantum advantage.\cite{Aspuru-Guzik2005,McArdle2018a,Cao2018}
While full error-correction is not expected to be realized in the near future, so-called Noisy Intermediate Scaled Quantum (NISQ) devices\cite{Preskill2018} have interesting properties that might still offer important computational advantages. 

While the first quantum algorithm proposed for simulating many-body systems, the Phase Estimation Algorithm (PEA),\cite{Kitaev1995,Lloyd1996,Aspuru-Guzik2005}
provides a path for achieving arbitrarily accurate simulations, it does so at the cost of incredibly deep circuits. Because device noise and errors limit the number of gates that can be applied in sequence, PEA is not viable on NISQ devices. In 2014, Peruzzo and coworkers proposed and demonstrated an alternative algorithm termed the  Variational Quantum Eigensolver (VQE)\cite{Peruzzo2014} which offers unique advantages for NISQ devices. Unlike PEA, VQE limits the depth of the circuit, which makes it possible to implement on current and near-term devices. However, this comes at the cost of an increased number of measurements, and the introduction of a wavefunction ansatz that can limit the accuracy of the simulation (although our recent approach,  ADAPT-VQE, can remove the ansatz error).\cite{adapt} 
The initial demonstration of VQE\cite{Peruzzo2014} was followed by several theoretical studies\cite{McClean2016,OMalley2016,McClean2017,Barkoutsos2018a,Romero2018,Colless2018,kUp} and demonstrations
on other hardware such as superconducting qubits\cite{OMalley2016,Kandala2017,Colless2018} and trapped ions.\cite{Shen2017,Hempel2018}

A key ingredient in VQE is the ansatz, which is implemented as a quantum circuit which constructs trial wavefunctions that are measured and then updated in a classical optimization loop. The quality of the ansatz ultimately determines the accuracy of the simulated ground state energy and properties. In the original proposal, the unitary variant of coupled-cluster theory was chosen as an ansatz due to several attractive features:
\begin{itemize}
    \item Accurate: Coupled-cluster theory is among the most accurate classical methods for many-body simulation.
    \item Well studied: The unitary variant of coupled-cluster singles and doubles (UCCSD) has been analyzed in detail in the context of classical simulations.\cite{bartlett_ucc_1989, kutzelnigg_ucc_1991, bartlett_ucc_2006, scuseria_ucc_2018}
    \item Unitary: Because a quantum circuit implements unitary operations, the unitary nature of UCCSD makes the approach natural in a VQE context.
\end{itemize}
The UCCSD ansatz is obtained by replacing the traditional Hermitian cluster operator terms in coupled cluster theory with anti-Hermitian operators:
\begin{align}
 \label{ucc}
|\Psi_\text{UCCSD}\rangle &=e^{\hat{\mathcal{T}}_1+\hat{\mathcal{T}}_2}|0\rangle \\
\hat{\mathcal{T}}_1&=\sum_{ia}\theta_{ia}\left(a^\dagger_aa_i-a^\dagger_ia_a\right)\nonumber\\
\hat{\mathcal{T}}_2&=\sum_{ijab}\theta_{ijab}\left(a^\dagger_aa^\dagger_ba_ia_j-a^\dagger_ja^\dagger_ia_ba_a\right),\nonumber
\end{align}
where $|0\rangle$ is the uncorrelated reference state, usually Hartree-Fock, $a^\dagger_p$ ($a_p$) is a creation (annihilation) operator for the orbital indexed by $p$, and $\{\theta_{ia}, \theta_{ijab}\}$ are the parameters to be variationally optimized.

Although the unitarity of UCCSD implies an ease of implementation on quantum hardware, gate-based quantum computing requires a decomposition of operations into one- and two-qubit gates, such as single-qubit rotations and CNOT gates. In contrast, complicating direct implementation, the $\hat{\mathcal{T}}_n$ operators simultaneously act on \textit{N} qubits. In principle any unitary operation can be decomposed into one- and two- qubit gates.\cite{nielsen_chuang_2010}
However, the number of gates produced from such a decomposition grows rapidly with the number of qubits acted on by the unitary, making it desirable to use an approximation scheme such as Suzuki-Trotter \cite{Hatano2005} when implementing \textit{N}-qubit unitary operators.

The first-order Suzuki-Trotter approximation is given by Eq. \ref{eq:approx}.
\begin{equation}\label{eq:approx}  
e^{\hat{A}+\hat{B}}\approx e^{\hat{A}}e^{\hat{B}}.
\end{equation}
This becomes exact in infinite order :
\begin{equation}\label{eq:exact}
e^{\hat{A}+\hat{B}}=\lim_{n\to\infty}\left(e^{\frac{\hat{A}}{n}}e^{\frac{\hat{B}}{n}}\right)^n.
\end{equation}
To approximate UCCSD accurately using a product form, large Trotter numbers, $n$, could in principle be used. This would, of course, create extremely deep circuits, making quantum simulation intractable. Alternatively, one could choose an aggressive truncation such as that in Eq. \ref{eq:approx}.  In general, this would provide a very poor approximation to the UCCSD wavefunction, but would provide a relatively shallow circuit that is better for NISQ realization. Note the stark difference between the effect of ``Trotterizing'' the ansatz in VQE, and Trotterizing the time-evolution operator for algorithms like PEA. 
In Trotterizing the evolution operator, the goal is to reproduce the dynamics of the original Hamiltonian. Any Trotter error destroys the dynamics and thus convergence with respect to Trotter error is sought.\cite{babbushChemicalBasisTrotterSuzuki2015,heylQuantumLocalizationBounds2019,siebererDigitalQuantumSimulation2019}
In contrast, when Trotterizing the ansatz in VQE it is generally accepted that the variational optimization can, in practice, absorb most of the energy difference between the conventional UCCSD and the Trotterized form.\cite{Barkoutsos2018a, Romero2018, rubin_hybrid_2016}

At this point we want to clarify some of the language used above.
Despite having used the Suzuki-Trotter approximation as a \textit{motivation} for separating out the ansatz into a product form, it is no longer appropriate to call this a Trotter approximation. The reason is that the Trotterization occurs before parameter optimization. Thus, one is actually variationally optimizing the parameters of the product form, and it no longer relates to the conventional UCCSD (in Ref. \citenum{evangelistaExactParameterizationFermionic2019} Evangelista et al. refer to this as the disentangled form of UCCSD, opting to avoid the Trotterization language altogether). In fact, if one were to use the optimized parameters from the product form, and insert them into the conventional UCCSD ansatz, the result would necessarily be higher in energy. Therefore, it is important to note that the term ``Trotterized form'' referred to throughout this letter is not an approximation to the conventional UCCSD ansatz. It is instead a different ansatz altogether, a point easily made by recognizing that the Trotterized form can sometimes yield a lower energy than the conventional, yet variational, UCCSD. 

Unfortunately, a problem of definition arises during Trotterization. Reordering the product approximation in Eq. \ref{eq:approx} does not generally give the same result, except in the trivial case where the operators commute. With the number of operator orderings being a path enumeration problem, the number of possible ans\"atze produced during Trotterization (and potentially reported in the literature) is exponentially large.  This of course is not an issue in UCCSD, as a sum of operators has no dependence on the order in which they are summed.  

 The objective of this letter is to determine if the term ``Trotterized UCCSD'' is sufficiently well-defined, 
such that the range of energies coming from different operator orderings falls within some notion of chemical accuracy (e.g., 1 kcal/mol), a term referred to in the title as \textit{chemically well-defined}.
If that were the case, then the term ``Trotterized UCCSD'' would be well-defined, as an arbitrary operator ordering would produce practically similar results. However, if changing the operator ordering significantly changes the accuracy on a chemical scale, then it proves necessary to provide more information to fully define an ansatz and to provide reproducible results. To answer this question, we perform classical simulations with randomly shuffled operators using a custom code built with OpenFermion\cite{openfermion} and Psi4,\cite{psi4} which uses the gradient algorithm we developed, which is outlined in the Appendix of Ref. \citenum{adapt}.
The results using various operator orderings are compared to both UCCSD and Full CI (FCI).

\section{Numerical Examples}
We consider four molecules in the context of the UCCSD ansatz, $\mathrm{LiH}$, $\mathrm{H_6}$, $\mathrm{BeH_2}$, and $\mathrm{N_2}$ with its $1s$ and $2s$ orbitals frozen. All molecules are arranged in uniform, linear geometries with varying interatomic distances. For each system, we classically simulate the calculation of a potential energy curve using a large number of random operator orderings.

\begin{center}
\begin{figure*}[ht!]
\centerline{\includegraphics[width = \linewidth]{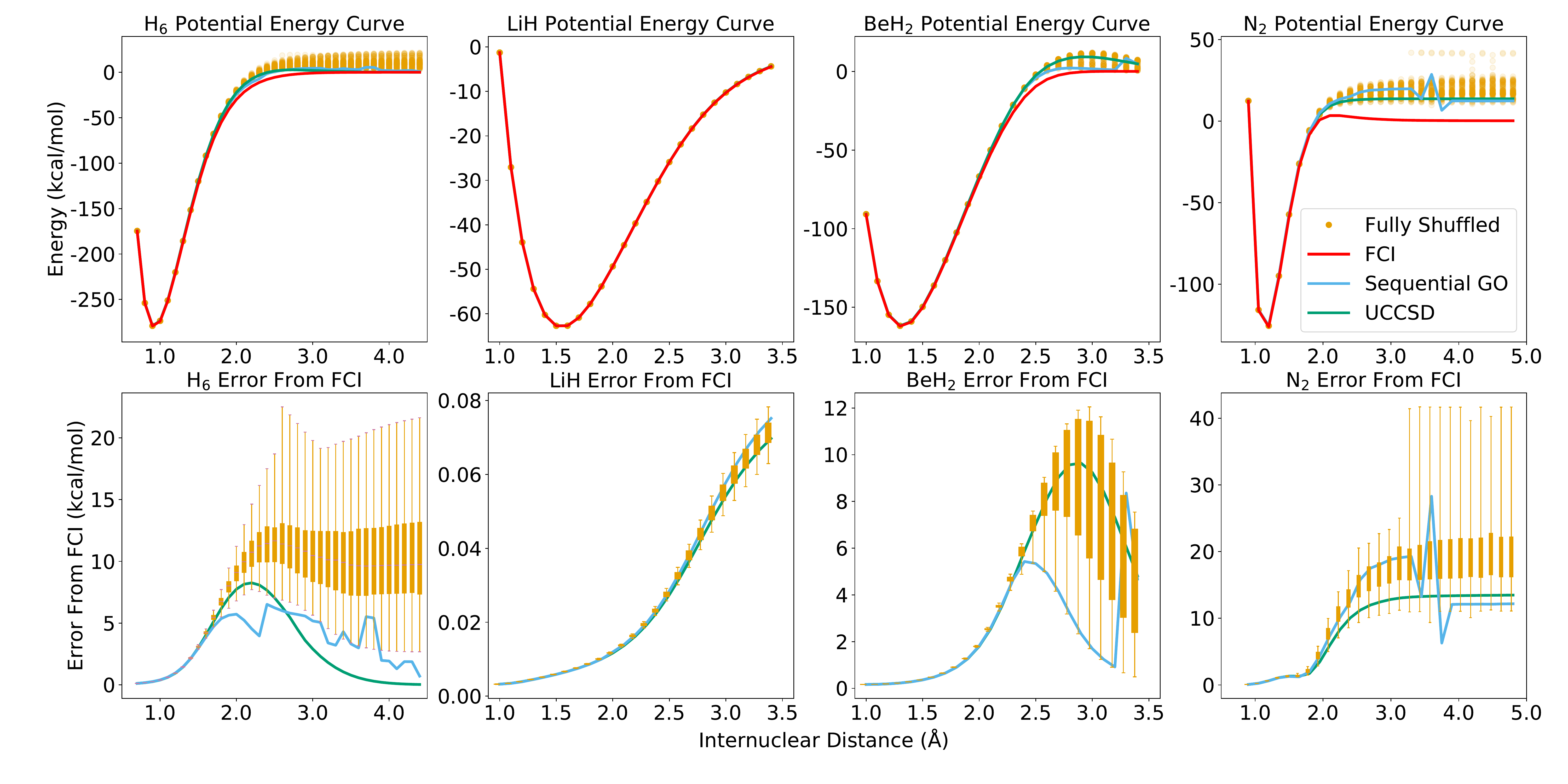}} 
\caption{\label{harper}Potential energy curves relative to the FCI dissociation limit of each system for, from left to right, $\mathrm{H_6}$, LiH, $\mathrm{BeH_2}$, $\mathrm{N_2}$ (top) and errors from FCI (bottom) for the UCCSD ansatz.}
\end{figure*}
\end{center}

For each system, the minimal STO-3G basis is used to minimize computational cost (the implementations use the full Hilbert space of the orbitals), and the restricted Hartree-Fock (RHF) singlet state is chosen as the reference state.  The one- and two- electron integrals are computed with the Psi4 quantum chemical package.\cite{psi4} The Hamiltonian, anti-Hermitian operators in the UCCSD ansatz, and reference state are formed in the qubit basis using the Jordan-Wigner transform in OpenFermion.\cite{openfermion}  At this point, the various orderings of ans{\"a}tze are constructed, and their parameters $\{\theta_{ia}, \theta_{ijab}\}$ are optimized by the SciPy implementation of BFGS.\cite{fletcher_practical_1987} The potential energy curves are displayed in Fig. \ref{harper} along with standard deviation plots and range plots.  We additionally compare the random Trotter orderings to a ``sequential gradient ordering,'' (SGO), a quasi-deterministic method where one operator with the largest gradient is added at a time, according to the prescription followed by the ADAPT-VQE ansatz construction.
However, in contrast to the ADAPT-VQE, the SGO approach refrains from allowing inclusion of more than one instance of the same operator so that a direct comparison to the original UCCSD results can be made. 

A cursory evaluation of the data suggests that the variance among different ans{\"a}tze increases with static correlation of the chemical system.  Because these tend to be the systems of greatest chemical interest for VQE since they represent classically hard problems, the ability to choose good Trotter orderings is critical.  

The UCCSD results for the first molecular PES, $\mathrm{H_6}$, are characterized by an accurate description near the equilibrium region, a quick increase in error upon bond breaking, and then a similarly rapid decrease in error as the bond is further stretched to dissociation. With five ``bonds'' being broken simultaneously, it is expected that UCCSD should fail to accurately describe this system. One interesting observation from this plot is that the \textit{ordering variance} (the statistical variance of the energies computed with randomly shuffled operators) increases as the UCCSD error increases. In contrast to H$_6$, LiH is a relatively simple system, and we observe negligible ordering variance. Regardless of Trotter ordering, the curves are all extremely good approximations.  

Similar to H$_6$, $\mathrm{BeH_2}$ exhibits a simultaneous quick rise in the ordering variance and UCCSD energy error. However, unlike H$_6$, the ordering variance decreases again after bond breaking, along with the UCCSD energy error. The range of values obtained from different orderings is of the same order of magnitude as the actual absolute error of the UCCSD energy.

Unlike both H$_6$ and BeH$_2$, the UCCSD curve for $\mathrm{N_2}$ does not decrease in error after bond breaking, but rather flattens out to a nearly constant error of around 10 kcal/mol. The ordering variance increases alongside the UCCSD error and also levels out, 
despite a significant jump occurring around 3.5 $\mathrm{\AA}$ in the range of energy values obtained from the Trotterized ans{\"a}tze. 
This is due to at least one of the operator orderings getting stuck in a local minimum (the variational parameters are initialized to 0), 
which is a consequence of the highly non-linear nature of the optimization. 

Overall, we find that when static correlation appears, the energy differences between orderings increases.  This can be understood from the fact that the differences between operator orderings depend on the commutators of the operators, and these in turn depend on the optimal parameter values, which tend to be larger when the electron correlation is stronger.  (A system with no electron correlation would have an optimal solution with all parameters equal to zero.)  As such it makes sense that for more strongly correlated systems, the differences between operator orderings increase.  While uniquely well-defined (up to orderings of operators with degenerate gradients), the sequential gradient ordering scheme does not appear to be reliably better or worse than other orderings.

\paragraph{Alternative ways to reorder operators}
In Fig. \ref{harper}, a comparison is made between the un-Trotterized ansatz and a series of randomly shuffled Trotterizations.  However, one could group the operators by excitation rank before Trotterization. This would result in a significantly reduced sampling space, and potentially provide more consistently accurate results. To address this possibility, we have computed the performance of multiple different orderings, such as grouping singles first and doubles second, or doubles first and singles second.  From these results, we find that it is generally favorable to apply double excitations to the reference first, followed by singles. This data is provided in the Supplementary Information. 

\paragraph{$k$-UpCCGSD}
From the results in Fig. \ref{harper}, we notice that the ordering variance increases with error in the associated un-Trotterized ansatz. It seems then that when UCCSD is accurate, there may be an excess of operators, such that the extra operators (while not necessary for accurate energy estimates) are useful in minimizing the differences between different Trotterization orderings. To test this hypothesis, we have additionally considered the more compact $k$-UpCCGSD ansatz by Lee et al.\cite{kUp}, which has far fewer parameters (for small $k$) than UCCSD, where $k$ controls the number of variational parameters by considering $k$ products of the ansatz with all generalized paired doubles and orbital rotations:

\begin{equation}
    \label{eq:kup}
    |\Psi_{k-\text{UpCCGSD}}\rangle = \prod_{i=1}^k\left(e^{\hat{T}^{(i)}-\hat{T}^{(i)\dagger}}\right)|0\rangle
\end{equation}
The $k$-UpCCGSD ansatz is a more economical parameterization where only the operators which are expected to be most important are included. This translates into having fewer excess parameters, such that higher accuracy can be reached with a comparable circuit depth by increasing $k$. Based on our results above, we would anticipate a higher ordering variance for small values of $k$ (larger than UCCSD), but that by increasing $k$, one can make the energy error (and thus the ordering variance) arbitrarily small. On the other hand, it is worth mentioning that the improvement attained by increasing $k$ is accompanied by placing a heavier burden on the classical optimizer, as it tends to exacerbate the highly non-linear character of the underlying optimization, making it difficult to locate the global minimum. Moreover, the minima found by the optimizer show strong dependence on the initialization of the variational parameters. One way this can be circumvented in the cases involving the un-Trotterized version of the $k$-UpCCGSD, as presented in Equation \ref{eq:kup}, is to perform many simulations with the variational parameters $\vec{\bm{\theta}}$ initialized at random, as suggested in Ref. \citenum{kUp}, and carried out here by repeating the simulations at each bond length 100 times and taking the lowest energy value as the global minimum for each geometry. This leads to potential energy curves for $k=$1, 2 that are smooth in the energy scale relevant in the current context. The variational parameters are initialized at 0 for all Trotterized ans\"{a}tze constructed based on Equation \ref{eq:kup}, in line with what is detailed for the UCCSD ansatz and whose results are displayed in Figure \ref{harper}.

\begin{figure}[ht!]
    \includegraphics[width=\columnwidth]{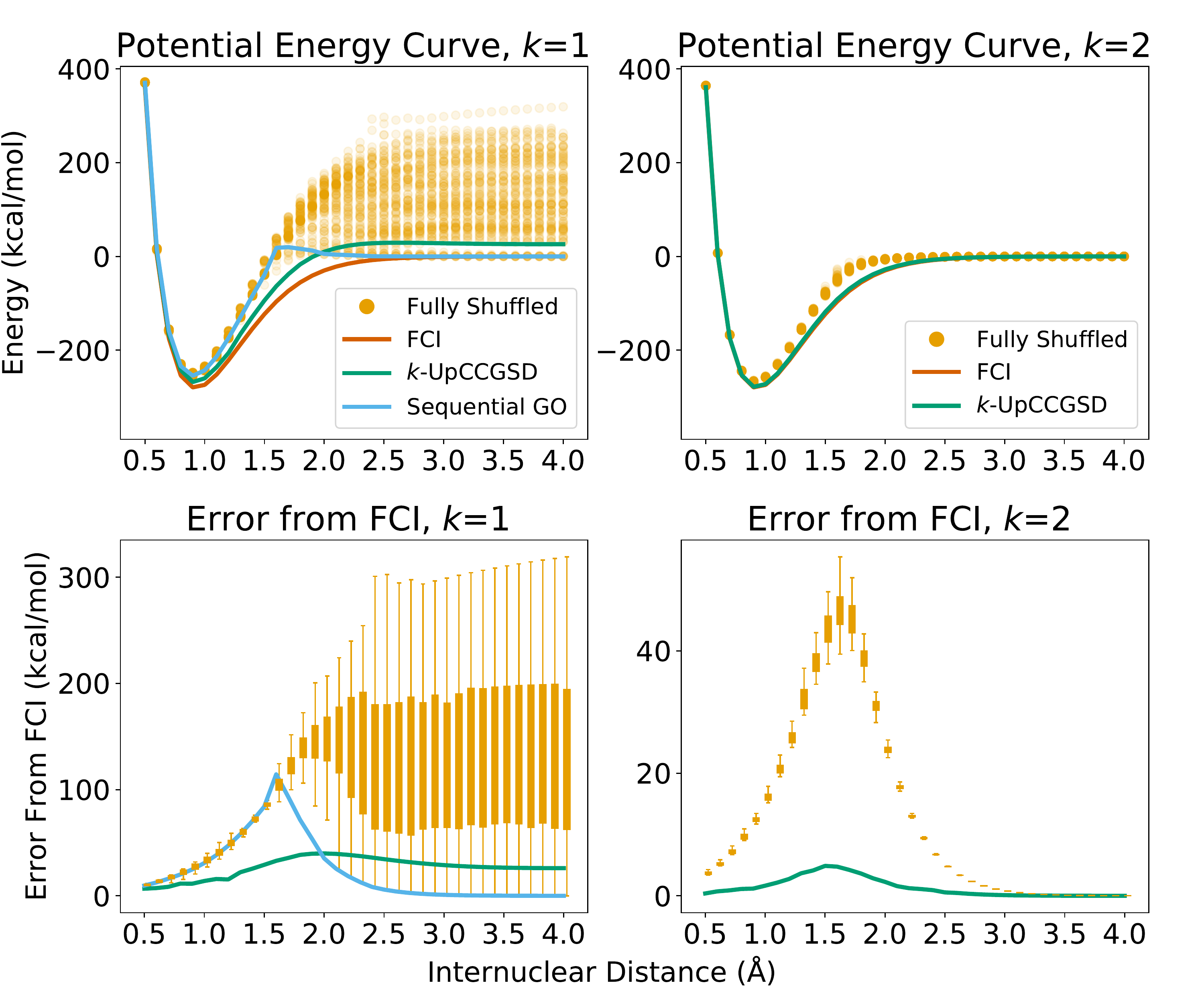}
    \caption{\label{kUp}Potential energy curves relative to the FCI dissociation limit of H$_6$ into six hydrogen atoms (top) and errors from FCI (bottom) for the 1-UpCCGSD (left) and 2-UpCCGSD (right) ans{\"a}tze.}
\end{figure}

Figure \ref{kUp} shows simulation results for H$_6$ with $k$=1, 2 for 100 randomly sampled operator groupings. Several features of the performance of the different Trotterized versions of 1- and 2-UpCCGSD agree with the results for the Trotterized versions of the UCCSD ansatz. For short bond distances ($<$1.1 \AA), there is an evident insensitivity of the energy with respect to a specific sampling of the operators. Despite being already fairly small in this regime with $k=1$, this distinction is largely quenched when $k=2$, rendering the results with differently sampled ans{\"a}tze visually identical on the scale of the plots.

The most remarkable divergences among the operator groupings and the size of the generator, that is, $k=1$ vs. $k=2$, are observed when moving toward the limit of H$_6$ dissociating into six non-interacting hydrogen atoms. These results are consistent with the observations found in Ref. \citenum{190912410v1Jastrowtype}, which noticed that for $k=1$ there were large differences in energy depending on whether one grouped or split the singles and doubles excitations. Ans{\"a}tze with different operator groupings start to deviate in the vicinity of the Coulson-Fischer point. In this region, none of the orderings that were sampled for the 1-UpCCGSD operators approach the corresponding un-Trotterized and FCI energies. Some of the ans{\"a}tze are able to get back on track in closely approaching the FCI dissociation limit, along with the un-Trotterized 1-UpCCGSD energies. These ans{\"a}tze happen to be largely comprised of double excitation operators flocked closer to the reference determinant, which is in line with the findings from the simulations with the SD orderings, provided in the Supporting Information. The 1-UpCCGSD ansatz tracks well the FCI results, being able to provide the correct qualitative behavior along the PES. However, this ansatz is quite compact, and its limited number of parameters impairs its ability to variationally achieve results that are quantitatively comparable to FCI. The operator ordering originated from the SGO construction closely follows the lowest energies from random operator samplings. It is worth pointing out that it is able to overcome the deficiencies around the Coulson-Fischer point and asymptotically recover the exact (FCI) dissociation limit, whereas the corresponding un-Trotterized ansatz cannot account for all the missing correlation as the H-H bonds are stretched. Because in the SGO ansatz the operators are added according to the magnitude of their gradient component, an ansatz with identical operators, such as $k$-UpCCGSD with $k>1$ cannot be unambiguously defined and that is why we do not report such results in Figure \ref{kUp}. 

The disparities among operator groupings are largely removed all throughout the potential energy curves by doubling the number of variational parameters, accomplished by setting $k=2$. We preserve the same orderings studied for $k=1$, that is, we have a product of two Trotterized exponential generators wherein operators are not shuffled across the two instances of $e^{\hat{T}^{(i)}-\hat{T}^{(i)\dagger}}$. Except for a slight spread surrounding the Coulson-Fischer point which is the region most strongly correlated in the potential energy curve, all of the different ans{\"a}tze behave in a strikingly similar fashion. The errors are largest in this region and, keeping in mind the different scales in the plots when changing $k$, they are significantly mitigated in comparison with $k=1$, with all orderings approaching the FCI energy in the dissociation limit. The advantage due to a larger set of variational parameters is also reflected in the un-Trotterized version of the ansatz, 2-UpCCGSD, whose dissociation curve practically overlays with the FCI results. The significant improvement in the results with $k=2$, accompanied by a virtually absent spread in the computed energies, is in agreement with the findings of Lee et al.,\cite{kUp} which implies that these ans\"atze are relatively insensitive to the ordering of the operators. 

\section{Conclusions}
In this letter, we sought to determine if the operator ordering in Trotterized UCCSD impacts the results in a `chemically meaningful' way, such that the differences between unique operator orderings produce results which differ on a chemical scale, i.e., greater than 1 kcal/mol. Our numerical simulations clearly demonstrate that the operator ordering has a significant effect (large energy differences between orderings) only when there is a significant amount of electron correlation. However, the renewed interest in UCCSD (and the relevance of the Trotterized form) is due to the use of the UCCSD ansatz in VQE simulations on quantum computers. Strongly correlated molecules are the primary target of quantum simulations, and so this makes the issue of operator ordering even more important. Consequently, the results in this paper emphasize that to ensure scientific reproducibility, it is necessary for authors to report the specific orderings used in simulations involving Trotterized ans{\"a}tze. These results strongly advocate for the  use of a dynamic ansatz which uniquely determines the operator ordering, such as ADAPT-VQE,\cite{adapt} or adopting an ansatz which does not require trotterization (such as the Jastrow-based approach in Ref. \citenum{190912410v1Jastrowtype}).   Our findings also suggest that there are systematic patterns to which Trotter orderings will give the lowest energy, offering a useful route to defining useful and unique operator orderings.

\section{Acknowledgements}
This research was supported by the U.S. Department of Energy (Award No. DE-SC0019199) and the National Science Foundation (Award No. 1839136). S.E.E. also acknowledges support from Award No. DE-SC0019318 from the U.S. Department of Energy.

\section{Supplementary Information}
A supplementary file is available which presents plots for simulations with operators shuffled only within excitation rank sets. Additionally, information is provided detailing how to access the code which was used to obtain these results. 

\providecommand*{\mcitethebibliography}{\thebibliography}
\csname @ifundefined\endcsname{endmcitethebibliography}
{\let\endmcitethebibliography\endthebibliography}{}

%\bibliography{paperNotes,adaptvqeRefs}
%\bibliographystyle{achemso}
\end{document}